# ENTROPY OSCILLATIONS

Viktor I. Shapovalov

The Volgograd Branch of Moscow Humanitarian-Economics Institute, Volgograd, Russia.
shavi@rol.ru

The opportunity of occurrence of entropy oscillations around of a stationary state in linear and nonlinear processes is theoretically shown. The new mechanism of global tendencies' appearance is described.

**1. Introduction**

To the present time within the framework of notion of entropy as a quantitative measure of the disorder there are two basic directions of use of the entropy approach for the description of processes of ordering in open systems. Let's allocate the main features of each of them.

1. According to [1],

$$\Delta S = \Delta_i S + \Delta_e S , \qquad (1)$$

where $\Delta S$ – entropy change of system; $\Delta_i S$ – entropy change, produced inside a system; $\Delta_e S$ – entropy outflow or inflow in system from the outside.

The given ratio allows allocating situations, in which entropy of open system can decrease. It is necessary to be $\Delta_e S < 0$ and $|\Delta_e S| > |\Delta_i S|$. However (1) has no indications of reasons defining sign and volume of $\Delta_e S$.

2. The inequality is known [2]:

$$S[X] \geq S[X | Y], \qquad (2)$$

where the square brackets are the designation, but not the functional dependence; $X$ and $Y$ – variables, determining the system state; $S[X | Y]$ – the conditional entropy of the stationary state after opening, as a result of which there were changes in the system described by the additional variable $Y$.

The inequality (2) has obvious defect. According to (2), the entropy cannot increase after opening. But if we apply this rule to the known task of the thermal contact of two bodies forming the isolated system, we shall come to the contradiction, because entropy is necessarily increased at least in one body after the contact (opening).

It was shown in papers [6-8], that this defect disappears**,** if we enter the entrostat concept.

By *the entrostat* we mean the external system, which doesn't change its own entropy while influencing the researched system. In practice the external environment in relation to the researched system is the entrostat meeting the following condition:

$$\frac{|\Delta S|}{S} \gg \frac{|\Delta S_e|}{S_e},$$

where $\Delta S$ and $\Delta S_e$ – the entropy changes of the researched system and external environment accordingly caused by their interaction.

As we see, the external environment is entrostat, if the entropy change of this environment can be neglected [3].

The main advantage of the entrostat concept introduction consists in fact that it enables to exclude the external environment in the behavior analysis of the open system.

The last make it possible to prove the following inequality for the systems under entrostat influence [4,6-8]

$$S[X] > S[X|Y_1] > S[X|Y_1Y_2] > ... > S[X|Y_1Y_2...Y_i] > ... > 0, \qquad (3)$$

where $S[X]$ – the entropy value of the closed system in the equilibrium state; $S[X|Y_1Y_2...Y_i]$ – the entropy value of the open system in the $i$-th stationary state, which differs from closed one by changes in the structure which has appeared due to the external influence and are described by variables $Y_1, Y_2, ..., Y_i$.

According to (3), the comparison of the open states of the system is correct, if it goes about the interaction of the system with *entrostat*. Neglecting of the specified circumstance results in the contradiction, this, in particular, was marked in the problem of the thermal contact of two bodies forming the isolated system. As it is impossible to consider any of the bodies as entrostat in the given task, the known inequality (2) cannot be applied.

The transitions from one inequality to another in expression (3) occur due to the change of volume of entrostat influence at the system. The entrostat influence is considered to be *constant* during some interval of time, sufficient for system to reach a stationary state.

The phenomenological parameter, which quantitatively describes the value of entrostat influence on the system, is named the *openness degree* $\alpha$ [3].

As it is visible from (3), each value of $\alpha$ an unequivocally corresponds to the certain stationary value of entropy. The limiting states of the system occupy the extreme positions of a line (3). $\alpha = 0$ is carried out for an extreme left position (absolutely closed state), for extreme right one: $\alpha = \alpha_{max}$ (it is maximum opened state):

$$\begin{array}{cccccc} S[X] > & S[X|Y_1] > & S[X|Y_1Y_2] > ... > & S[X|Y_1Y_2...Y_i] > ... > 0 \\ \alpha = 0 & \alpha_1 & \alpha_2 & \alpha_i & \alpha_{max} \\ S_{AC} & S_{\alpha_1} & S_{\alpha_2} & S_{\alpha_i} & \end{array}$$

Here: $S_{AC}$ – entropy of an equilibrium state of system in absence of external influence (the system is absolutely closed); $S_{\alpha_i}$ – entropy of a stationary state of system having a degree of an openness $\alpha_i$.

Inequality (3) can be graphically represented as entropy line in a figure 1. At this figure $S_0$ – is initial value of entropy. The blacked out part of column shows level entropy of system in a stationary condition.

It was also shown in [3,4,6-8], that

$$\Delta S_\alpha = \Delta S_{AC} + \Delta S_\alpha^-, \qquad (4)$$

where (see figure 1): $\Delta S_\alpha = S_\alpha - S_0$ – entropy change of the system having a degree of an openness $\alpha$ and which has reached a stationary state; $\Delta S_{AC} = S_{AC} - S_0 > 0$ – entropy change of the absolutely closed system which has reached balance;

$$\Delta S_\alpha^- = S_\alpha - S_{AC} < 0 \qquad (5)$$

– negative entropy change, on which the stationary value $S_\alpha$ decreases with increase of $\alpha$.

The equality (4) differs (1) by the fact, that $\Delta S_\alpha^-$ is always less zero.

According to (3) (see also figure 1), the value of $\Delta S_\alpha^-$ unequivocally corresponds to the openness degree of the system:

$$\alpha_i \Leftrightarrow \Delta S_{\alpha_i}^-. \tag{6}$$

$\Delta S_\alpha^-$ was named in our papers [3-8] as *a critical level of the system organization*. We examine $\Delta S_\alpha^-$ as a quantitative measure of system organizing in a stationary state. According to (3) under the organization of system we shall understand the structural links described by macroscopic variables $X, Y_1, Y_2, ..., Y_i, ...$

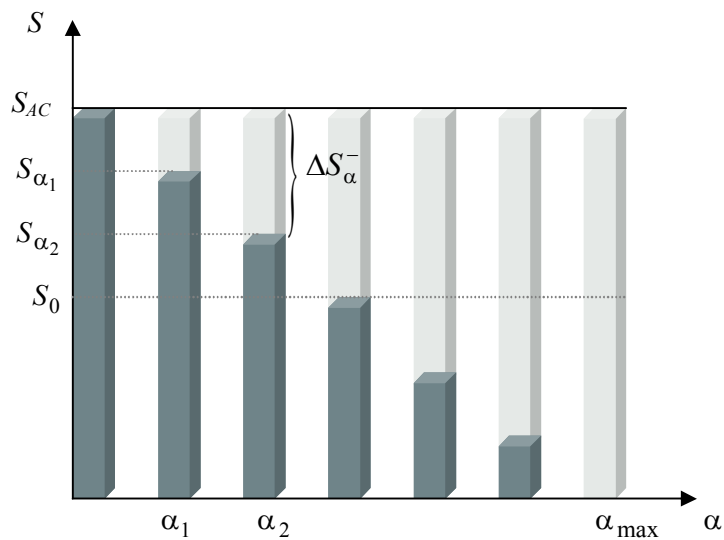

Figure 1. *The entropy line of system* – values of entropy $S$ for stationary states of system at its various degrees of openness $\alpha$ (from $\alpha = 0$ up to $\alpha_{max}$) [3,4].

Two important conclusions follow from (6) [3,7,8]

1. If the system is organized above its critical level $\Delta S_\alpha^-$, then the processes increasing entropy prevail in it, if lower – the processes reducing entropy are prevailing (at the critical level itself the action of the specified processes compensate each other, and the state of the system becomes stationary).

2. To increase or to reduce the critical level value $\Delta S_\alpha^-$ it is necessary accordingly to increase or to reduce the openness degree of $\alpha$.

In the following sections we use these conclusions for back grounding of entropy oscillations phenomena.

**2. Entropy oscillations in linear processes.**

Let us write some known equations.

$$\sigma = \frac{ds}{dt} = -divJ + \sigma_i = \sigma_e + \sigma_i, \tag{7}$$

where $\sigma$ – rate of the local entropy change; $s = dS/dV$ – local entropy; $S$ – entropy of the system; $V$ – volume; $t$ – time; $J$ – density of an entropy flow; $\sigma_i$ – production of the local entropy (part of rate of local entropy change $\sigma$, caused by internal processes).

$$P = \int \sigma \, dV = dS/dt \qquad (8)$$

– the rate of system entropy change. According to (7), $P = P_e + P_i$, where $P_i = \int \sigma_i \, dV$ – the entropy production in the system (a part of rate of entropy change in system, caused by internal processes).

$$\frac{\partial P_i}{\partial t} \leq 0 \qquad (9)$$

– the theorem of the minimal entropy production [1]. It was shown in [7,8], that for systems under entrostat influence, inequality (9) transforms to

$$\frac{\partial |P|}{\partial t} \leq 0, \qquad (10)$$

where $P$ defines from (8). Inequality (10) as well, as (9) is correct for linear processes. In the process of (10) proving, the conclusions given at the end of section 1 were taken into consideration.

We shall add some function $F(P,S)$ to the left part of an inequality (10) in order to transform an inequality in equation:

$$\frac{\partial P}{\partial t} + F(P,S) = 0 \qquad (11)$$

(we have omitted a sign of the module, as for the equation it has no basic importance). Let's notice that the function $F$, added in (11) to $\partial P/\partial t$, should have the same sense, as $\partial P/\partial t$. Therefore in general case it is necessary to write down: $F = F(\partial P/\partial t, P, S)$. But in linear approach the account of $\partial P/\partial t$ as function $F$ variable does not result essential change of the received below equation of entropy oscillations and it is reduced only to remarking of constant factors.

As (10) is correct for linear processes, let us present $F$ as

$$F(P,S) = F(P_\alpha, S_\alpha) + \beta(P - P_\alpha) + \eta(S - S_\alpha) = \beta P + \eta S - \eta S_\alpha;$$
$$\beta = (\partial F/\partial P)_{P_\alpha}; \quad \eta = (\partial F/\partial S)_{S_\alpha}.$$

The index "$\alpha$" specifies that value of term is undertaken in a stationary state, the degree of which openness is equal to $\alpha$. Besides, it was taken into account that $F(P_\alpha, S_\alpha) = 0$ and $P_\alpha = 0$ (as $F$ and $P$ are equal to zero in a stationary state). In result (11) will get a form of the equation of oscillations [7,8]:

$$\frac{\partial^2 S}{\partial t^2} + \beta \frac{\partial S}{\partial t} + \eta S = \eta S_\alpha. \qquad (12)$$

If we express $S_\alpha$ from (5) and substitute it in (12), then we shall receive entropy oscillations around of a critical level of organization of system $\Delta S_\alpha^-$. The simple form of such oscillations for a case $\eta > \beta^2/4$ is given in a figure 2.

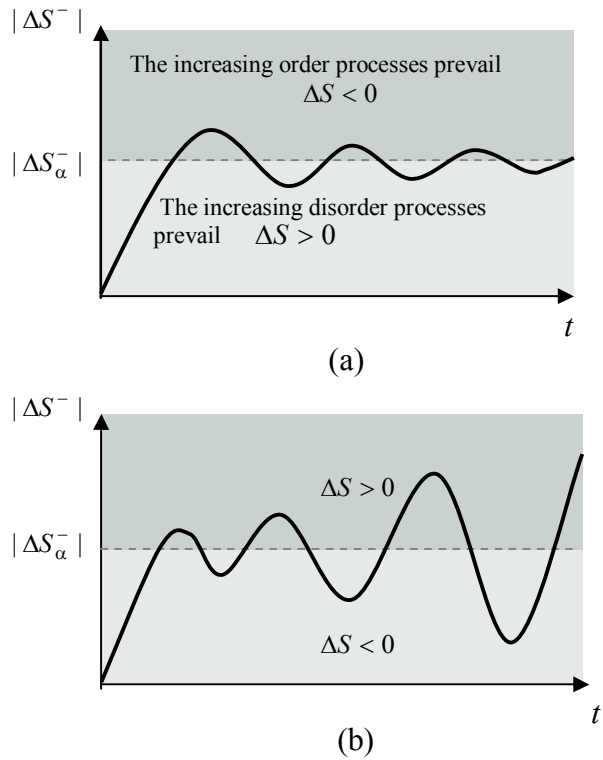

Figure 2. The entropy oscillations around a stationary state with a degree of an openness $\alpha$ [7]. Here: $|\Delta S^-| = |S - S_{AC}|$; (a): $\beta > 0$; (b): $\beta < 0$.

### 3. Entropy oscillations in nonlinear processes

Let us write the known expression:

$$\sigma = \sum X_k I_k; \quad X_i = \partial s/\partial a_i, \quad I_i = da_i/dt \qquad (13)$$

где $a_i$ – is some parameter of the system; $\sigma = ds/dt$ (see (7)); $s = dS/dV$.

We shall understand the process in which $I_i(X_i)$ is nonlinear function as nonlinear process.

Let us write the Gibbs formula for the system consisting of some components ($\gamma = 1, 2, ..., n$):

$$T\, d\hat{s}/dt = de/dt + p\, dv/dt - \sum_\gamma (\mu_\gamma\, dN_\gamma/dt), \qquad (14)$$

where $T$ – temperature; $\hat{s} = \rho^{-1}\, dS/dV$ – specific entropy; $\rho$ – density; $e = \rho^{-1}\, dE/dV$; $E$ – internal energy; $p$ – pressure; $v = \rho^{-1}$ – specific volume; $\mu_\gamma$ – specific chemical potential of component ; $N_\gamma$ – component concentration ($\sum_\gamma N_\gamma = 1$, see [1]).

Let us use the equation of uninterruptness:

$$\frac{\partial \rho}{\partial t} + \frac{\partial(\rho \upsilon_j)}{\partial x_j} = 0, \qquad (15)$$

where $x_j$ – coordinate; $\upsilon_j = dx_j/dt$. It's easy to see, that

$$\rho \frac{d\hat{s}}{dt} = \frac{\partial(\rho \hat{s})}{\partial t} + \frac{\partial(\rho \hat{s} \upsilon_j)}{\partial x_j},$$

Let system be motionless and convection is absent in it: $\upsilon_j = 0$, $d/dt = \partial/\partial t$. Then

$$\rho \frac{d\hat{s}}{dt} = \frac{\partial(\rho \hat{s})}{\partial t} = \sigma,$$

and (14) transforms to:

$$\sigma = \rho \left( T^{-1} \frac{\partial e}{\partial t} - \sum_\gamma \mu_\gamma T^{-1} \frac{\partial N_\gamma}{\partial t} \right).$$

Here we take into account, that $\partial v/\partial t = \rho^{-2} \partial \rho/\partial t = 0$ (see (15) at $\upsilon_j = 0$).

With the help of (13) and reasoning, adduced in [1], we find:

$$\rho^{-1} \frac{\partial \sigma}{\partial t} = \underbrace{\frac{\partial T^{-1}}{\partial t} \frac{\partial e}{\partial t} - \sum_\gamma \frac{\partial(\mu_\gamma T^{-1})}{\partial t} \frac{\partial N_\gamma}{\partial t}}_{\rho^{-1} \sum \frac{\partial X_k}{\partial t} J_k} + \underbrace{T^{-1} \frac{\partial}{\partial t}\left(\frac{\partial e}{\partial t}\right) - \sum_\gamma (\mu_\gamma T^{-1}) \frac{\partial}{\partial t}\left(\frac{\partial N_\gamma}{\partial t}\right)}_{\rho^{-1} \sum X_k \frac{\partial J_k}{\partial t}} =$$

$$= -T^{-1} \left[ c_v T^{-1} \left(\frac{\partial T}{\partial t}\right)^2 + \sum_{\gamma\beta} \mu_{\gamma\beta} \frac{\partial N_\gamma}{\partial t} \frac{\partial N_\beta}{\partial t} \right] + \rho^{-1} \sum X_k \frac{\partial J_k}{\partial t} = \rho^{-1} \frac{\partial_X \sigma}{\partial t} + \rho^{-1} \frac{\partial_J \sigma}{\partial t},$$

where $c_v$ – a specific heat; $\mu_{\gamma\beta} = (\partial \mu_\gamma / \partial N_\beta)_{T,p}$ (see [1]);

$$\frac{\partial_X \sigma}{\partial t} = -\frac{\rho}{T} \left[ \frac{c_v}{T} \left(\frac{\partial T}{\partial t}\right)^2 + \sum_{\gamma\beta} \mu_{\gamma\beta} \frac{\partial N_\gamma}{\partial t} \frac{\partial N_\beta}{\partial t} \right].$$

Let us use (8):

$$\frac{\partial_X P}{\partial t} = \int \frac{\partial_X \sigma}{\partial t} dV = -\int \frac{\rho}{T} \left[ \frac{c_v}{T} \left(\frac{\partial T}{\partial t}\right)^2 + \sum_{\gamma\beta} \mu_{\gamma\beta} \frac{\partial N_\gamma}{\partial t} \frac{\partial N_\beta}{\partial t} \right] dV. \qquad (16)$$

Let us note, that the same calculations were made in [1] for entropy production $P_i$, which is the part of $P$ (see (8)).

In (16) the expression in square brackets is the square form. Hence, the sign $\partial_X P/\partial t$ depends only on a signs of $c_v$ and $\mu_{\gamma\beta}$.

For the system which are being under entrostat influence, it is possible to take advantage of the conclusions resulted in the end of section 1. If the system is organized below the critical level $\Delta S_\alpha^-$ then the processes reducing entropy prevail in it ($dS < 0$, $P < 0$). In this case $\partial_X P/\partial t > 0$. Hence, $c_v < 0$ and $\mu_{\gamma\beta} < 0$. Such system is unstable and can develop to a new state.

If the system is organized above $\Delta S_\alpha^-$ then the processes increasing entropy prevail in it ($dS > 0, P > 0$). In this case $\partial_X P/\partial t < 0$, $c_v > 0$, $\mu_{\gamma\beta} > 0$. Hence, the system is stable and comes back to an initial state.

It is easy to generalize both cases in the form of an inequality

$$\frac{\partial_X |P|}{\partial t} \leq 0. \tag{17}$$

Let's notice, that if one replace $P$ for $P_i$ at $c_v > 0$ and $\mu_{\gamma\beta} > 0$ an inequality (17) will coincide with known Glansdorff- Prigogine inequality [1].

Further we shall act in the same way as in section 2. We shall add some function $F(P,S)$ to the left part of an inequality (17):

$$\frac{\partial_X P}{\partial t} + F(P,S) = 0. \tag{18}$$

$$F(P,S) = F(P_\alpha, S_\alpha) + \beta_1(P - P_\alpha) + \beta_2(P - P_\alpha)^2 + \ldots + \eta_1(S - S_\alpha) + \eta_2(S - S_\alpha)^2 + \ldots,$$

$$\beta_k = \frac{1}{k!}\left(\frac{\partial^k F}{\partial P^k}\right)_{P_\alpha}; \quad \eta_k = \frac{1}{k!}\left(\frac{\partial^k F}{\partial S^k}\right)_{S_\alpha}.$$

Near to a stationary state $P - P_\alpha$ and $S - S_\alpha$ are small values. Hence, $F(P,S) \approx \beta_1 P + \eta_1 S - \eta_1 S_\alpha$ (we have considered that $F(P_\alpha, S_\alpha) = 0$ and $P_\alpha = 0$). As a result the equation (18) will be transformed to the equation of entropy oscillations in the form of (12). The unique difference is that the reason of these oscillations occurrence for nonlinear processes is change of the generalized forces $X_i$ at the constant generalized flows $I_i$.

### 4. Instead of the conclusion: The formation of the global tendencies

So, we have shown that near to stationary state entropy oscillations can arise. We shall notice that this conclusion is proved only for the systems which are being under influence entrostat.

In opinion of the author, in practice, entropy oscillations should be shown in the form of alternation of processes of self-organization and processes of disorganization. These oscillations can arise after the system will reach the critical level of the organization (see figure 2). The last has formed the basis for creation of a hypothesis about the reasons of formation of global tendencies. In opinion of authors of works [3,5,7,9], these tendencies may be responsible for the global problems which have arisen before mankind recently.

While transforming the contemporary world the mankind increases or reduces the order in it, i.e. changes the entropy of its habitat. There arises a question, what is the final entropy change, produced by the whole mankind: is it more or less than zero? The laws described in the present work, enable to receive the certain answer.

For the "Earth" system its rather constant openness degree in relation to the space (entrostat) sets the certain critical level of organization. The processes of ordering and self-organization ought to prevail on the Earth below the critical level ($\Delta S < 0$), and the processes of disorganization ($\Delta S > 0$) should have a higher critical level – see figure 2. In the first case the mankind, transforming the contemporary world, increases the order in it more, than disorder as a whole. Till when can it proceed? So long while creating it will not exceed the critical level of the planet organization $\Delta S_\alpha^-$. In this case the processes of disorganization will prevail. As a result the probability of destructive events will be raised, and the surplus, which the mankind has constructed, exceeding a critical level, will be destroyed (or will be compensated by destructions in the environment). By inertia it will be destroyed a little bit more, than it is necessary to be lowered up to a critical level. Below a critical level the processes of self-organization will prevail, and the mankind will again build houses and factories, partition off rivers by dams and etc., i.e. to reduce entropy of the Earth. After some time it will again exceed a critical level. Then everything repeats again – the entropy oscillations arise [3].

During the period of critical level exceeding the entropy rules raise the probability of any events contributing to the increasing of the disorder at the planet, i.e. form the destructive tendency [7,9]. What can become the evidence of the given tendency revealing? Firstly, the increase of calamities' intensity, disintegration of ecosystems, destructive fluctuations of the climate. Secondly, the probability increase of technogenic catastrophes, accidents and military conflicts. And thirdly, – any other destructive phenomena.

Pay your attention: all these three classes of the phenomena have a common denominator – the increase of the disorder. According to the author of the present article, the latter means, that these phenomena get in sphere of action of the laws described in the present work.

The understanding of the reasons of the given tendency not only enables to foresee future hardships, but also prompts a way to avoid them.

Let's remind (see section 1), that the value of the critical level of organization $\Delta S_\alpha^-$ unequivocally depends on the openness degree of $\alpha$. If we increase $\alpha$, we shall increase $\Delta S_\alpha^-$. This action will result in prevailing ordering processes and self-organization processes in the system. As a result the tendencies having creative character will be generated in it (i.e. the probability of events contributing to creation will be raised), but the destructive tendencies will gradually disappear.

Thus, to have the prevailing processes of ordering and self-organization in system "Earth" it is necessary to increase the value of its critical level of organization. For this purpose it is necessary to increase its openness degree (for example, as a result of purposeful and large scale intervention into the space). How could it be shown in practice? Here is a very evident example: constructing of industrial objects of the Moon – the nearest proportional to the Earth object – can be carried out only due to the joint efforts of many states. Hence, instead of war nations ought to be united. Then it would be the turn of the Mars and etc. Thus each time, as soon as the mankind would linger over the next scale opening, the threat of common destruction would arise again.

Let's get a brief result. The rather constant degree of an openness of the Earth sets the certain critical level of order at the planet. The mankind, creating in peacetime, inevitably aspires to exceed this level. And when it occurs, then, according to the author, the processes of disorganization should necessarily prevail at the Earth (i.e. the probability of any events conducting to destructions ought to raise). At the same time increase of an openness of a planet would also increase value of its critical level of organization that would result in formation of the steady creative tendencies [3,7,9].